\documentclass[aps,twocolumn,superscriptaddress,showpacs]{revtex4}
\usepackage{graphicx}
\begin{document}
\draft
\title{\bf  Spiral versus modulated collinear phases in the quantum ANNNH model}
\author{ J. Oitmaa }
\affiliation{School of Physics, The University of New South Wales,
 Sydney, NSW 2052, Australia}
\author{ R.R.P. Singh }
\affiliation{Department of Physics, University of California,
Davis, CA 95616, USA}
\date{\today}
\begin{abstract}
Motivated by the discovery of spiral and modulated collinear phases in several magnetic materials,
we investigate the magnetic properties of Heisenberg spin S=1/2 
antiferromagnets in 2 and 3 dimensions, with frustration arising
from 2nd-neighbor couplings in one axial direction (the ANNNH model).
Our results clearly demonstrate the presence of an incommensurate
spiral phase at T=0 in 2 dimensions, extending to finite temperatures
in 3 dimensions. The crossover between N\'eel and spiral order occurs
at a value of the frustration parameter considerably above the classical
value 0.25, a sign of substantial quantum fluctuations. We also 
investigate a possible modulated collinear phase with a wavelength of 4 lattice
spacings, and find that it has substantially higher energy and hence is not realized in the model.
\end{abstract}

\pacs{PACS Indices: 75.10.-b,75.10.Jm,75.30.Kz \\
\\  \\
(Submitted to  Phys. Rev. B) }
\maketitle
\newpage

\section{INTRODUCTION}
\label{sec1}

Frustrated magnetic materials continue to provide a fruitful interaction 
between experiment and theory [1]. In particular, large quantum fluctuations
 in systems with low spin and low dimensionality, coupled with frustration,
can lead to novel states [2], quite different from the usual N\'eel state
of classical antiferromagnetism.

Magnetic frustration can arise from the lattice structure itself, as in 
triangular and kagome systems, or from the presence of additional further-
neighbor interactions which favor a different type of order from that which
would arise from nearest-neighbor interactions alone. One such scenario is the 
inclusion of 2nd-neighbor interactions along one axial direction, in square 
or cubic lattices, referred to as the ANNNH (axial next-nearest-
neighbor Heisenberg) model [3-5].

The ANNNH model is the obvious quantum extension of the Ising version, the ANNNI
model, which was much studied primarily in connection with
modulated phases in both magnetic and alloy systems [6,7]. The ANNNI model was
found to have an extremely rich finite temperature phase diagram, in both
2 and 3 dimensions, with modulated phases having both constant and continuously
varying wavevectors.

Our motivation for studying the quantum ANNNH model is twofold. Firstly,
there are now a number of materials where commensurate-incommensurate transitions
and modulated spiral and collinear phases have been recently observed to arise  [8-11].
For example, in the materials Lu$_{1-x}$Sr$_x$MnSi, cycloidal antiferromagnetic
order is argued to arise from an axial next-neighbor interaction [9]. 
In the material BiMn$_2$PO$_6$ also a number of commensurate and incommensurate
phases are observed, driven by the spatial anisotropy of the interactions in
a 3-dimensional spin system [10]. On the other hand, the material FeSe shows a 
`pair-checkerboard' collinear magnetic order [11]. It would be intersting to establish
if such modulated collinear phases also arise in ANNNH models, like in their Ising counterpart.

Secondly, on general grounds one expects that the
presence of further neighbor interactions will favour spiral phases, in which
the average moment varies sinusoidally with a wavevector along the frustration
axis. While it is easy to demonstrate this for classical vector spins, we are 
not aware of any rigorous demonstration of this in the extreme quantum case 
S=1/2. It is well known that quantum fluctuations can stabilize collinear phases [16].
Thus, it is interesting to ask if such modulated collinear phases are stabilized in these systems
due to quantum fluctuations.

We find that such spiral phases do indeed  arise in the quantum models. In two-dimensions,
long-range order only arises at zero temperature, but in 3-dimensional systems,
such phases extend to finite temperatures, and their is a Lifhitz point where
Neel, spiral and paramagnetic phases meet [15]. We find that, despite strong quantum fluctuations,
the ANNNH model does not support modulated collinear phases. Instead, the parameter region
for the stability of the collinear Neel phase is substantially enhanced by quantum fluctuations.

We consider a Heisenberg model with Hamiltonian

\begin{equation}
H = J_0 \sum_{<ij>}^{(0)} {\bf S_i \cdot S_j} + J_1 \sum_{<ik>}^{(1)}
 {\bf S_i \cdot S_k} + J_2 \sum_{<il>}^{(2)} {\bf S_i \cdot S_l}
\end{equation}

\noindent where the sums are over nearest-neighbor bonds perpendicular to the
modulation axis, nearest-neighbor bonds along the modulation axis, and next-
nearest pairs along the modulation axis, with coupling constants $J_0, J_1, J_2$
respectively. This is shown in Fig. 1 for the 2-dimensional (2D) case.
The $\bf S_i$ are quantum spin S=1/2 operators. In the present work we consider
all interactions to be antiferromagnetic ($J_i > 0$), although other cases 
could be treated in a similar way.

\begin{figure}
 \includegraphics[width=8.0cm]{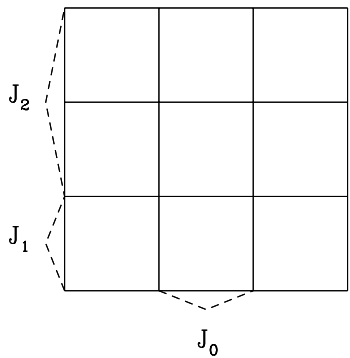}
 \caption{Coupling constants of the ANNNH model in two-dimensions}  
 \label{fig1}
\end{figure}

The phase diagram for classical spins is well known, but we repeat the argument
here for completeness. The energy of a classical spiral ground state is

\begin{equation}
    {E\over NS^2} = -nJ_0 + J_1 \cos q + J_2 \cos 2q
\end{equation}
 
\noindent where q is the angle between neighboring spins in the modulation
direction and n=1(2) for the square(simple cubic) lattice. Minimization
gives

\begin{eqnarray}
  q & = & \pi       \qquad \qquad \qquad \qquad J_2/J_1<1/4 \nonumber \\
    & = & \pi - \cos^{-1} ({J_1\over 4 J_2})   \qquad J_2/J_1>1/4.
\end{eqnarray}

Thus the small $J_2$ N\'eel phase becomes an incommensurate spiral with
wavevector q at the transition point $J_2=J_1/4$. It can also be seen 
that in the large $J_2$ limit, where $q \rightarrow \pi/2$, a collinear phase
in which each column has two spins 'up' followed by two spins 'down', with 
neighboring columns ordered antiferromagnetically, will become degenerate
with the spiral. Such a phase has been termed [9] 'pair-checkerboard',
but we will refer to it as a '2+2 phase'. Such a phase occurs in the ANNNI
model for large frustration and, while in the classical vector case (Eqn.3)
it only occurs as a limiting case, its stabilty in the quantum case has 
not been investigated previously, to our knowledge.

\begin{figure}
 \includegraphics[width=2.5cm]{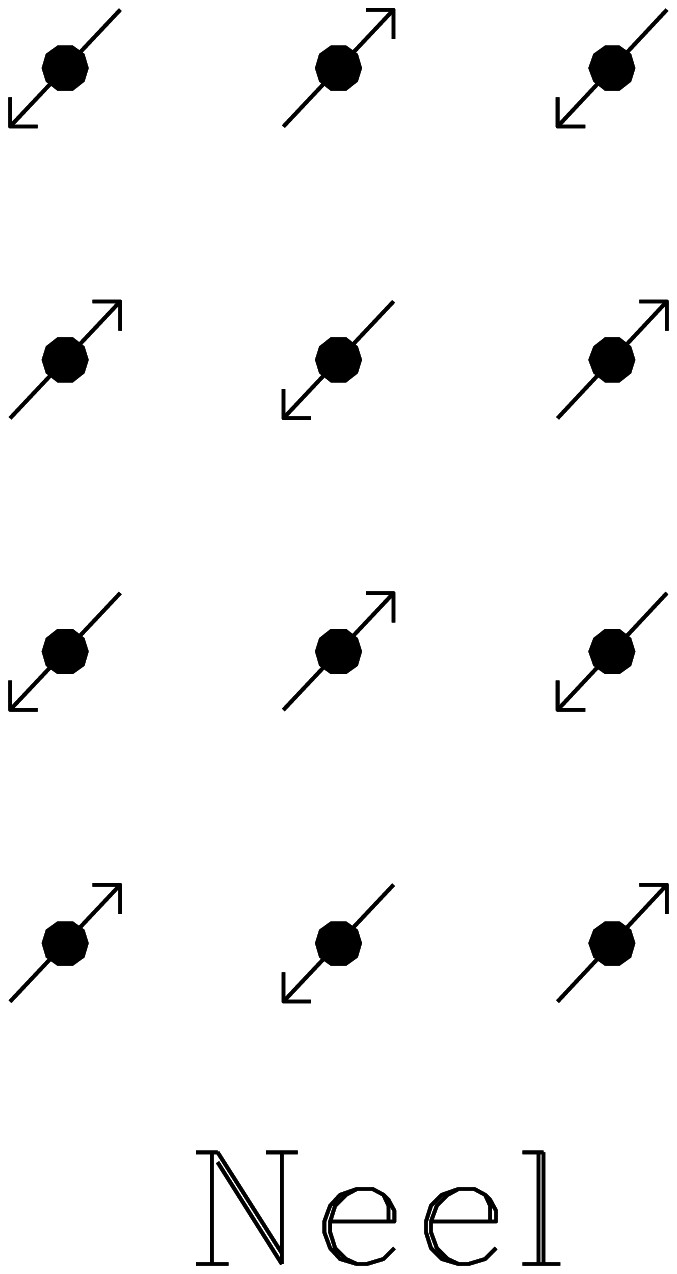}
 \hspace{0.4cm}
 \includegraphics[width=2.5cm]{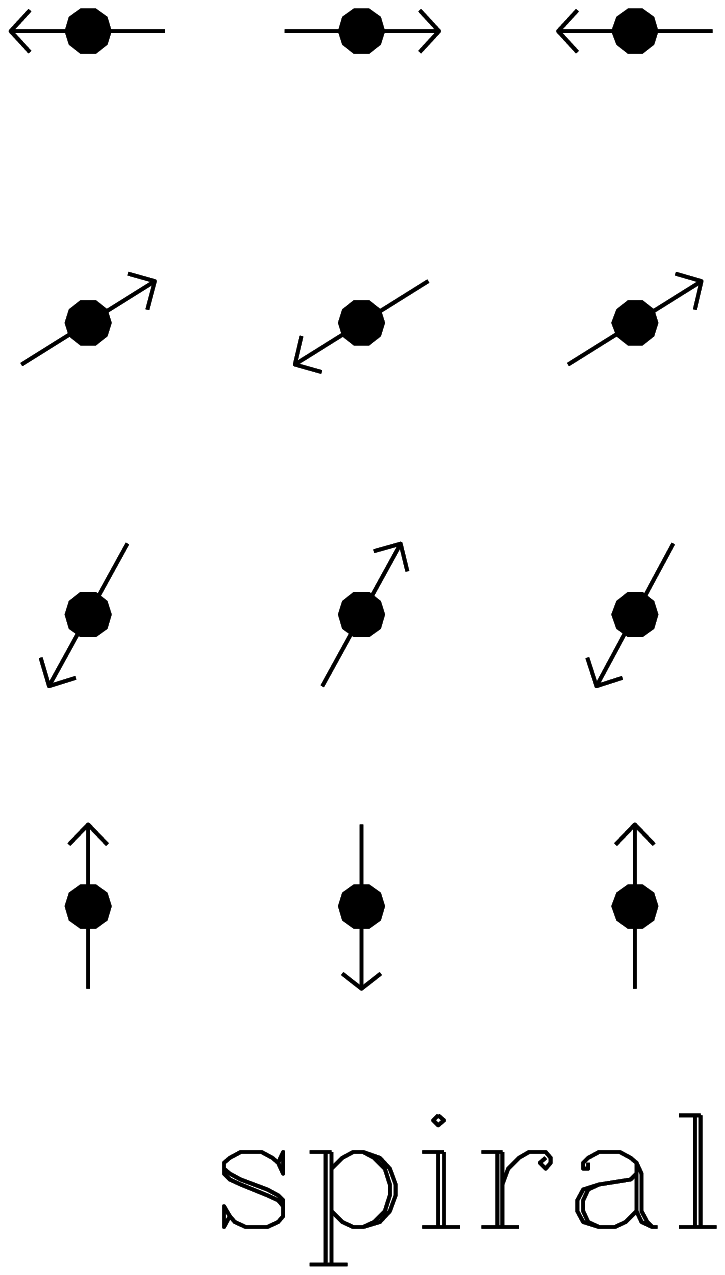}
 \hspace{0.4cm}
 \includegraphics[width=2.5cm]{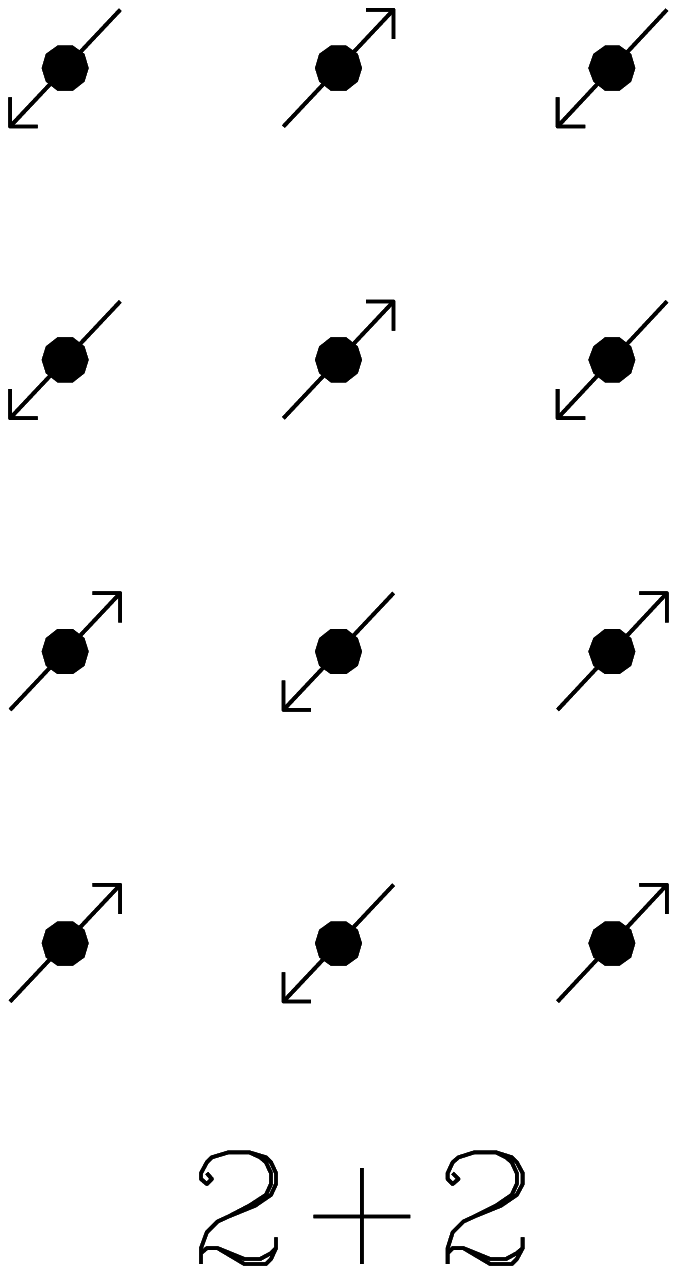}
 \caption{The N\'eel, spiral and 2+2 states}
 \label{fig2}
\end{figure}

A number of studies of the quantum ANNNH model were reported in the 
1980's [3-5], using bosonic Hamiltonians obtained via standard Holstein-
Primakoff or Dyson-Maleev transformations. These studies, which 
focussed only on the case of ferromagnetic $J_0 , J_1$, encountered
difficulties in treating quantum corrections about the classical states
in a consistent way. In any case, these analytic approaches are essentially
large S theories, and their reliability for S=1/2 is uncertain. We are not 
aware of any previous work on this model with antiferromagnetic first-
neighbor interactions.

Our aim in the present work is to explore the physics of this model for
spin 1/2, using series expansion methods [12,13]. This approach has been amply
demonstrated to give reliable results for quantum spin models, and is
a method of choice for models with strong frustration, where Quantum
Monte Carlo methods are defeated by the infamous 'minus sign' problem.
In the following sections we derive and analyse series for the ground
state energy and magnetization for both the 2D and 3D models. In Section
4 we compute series at high T for spin-spin correlations and for the
structure factor S(q). This analysis clearly shows that the large $J_2$
phase is an incommensurate spiral. Following the 2D work, in Section 4
we treat the 3D model, and present results at both T=0 and high T. Clear
differences from the 2D case are demonstated. Finally, in Section 5, we
summarize our results and suggest possible extensions of this work.

\section{GROUND STATE OF THE 2D ANNNH MODEL}
\label{sec2}
We use the linked-cluster method [12,13] to obtain series for the ground 
state energy and magnetization. In this approach, series are computed for 
a sequence of finite connected clusters, and these are combined to obtain
series in the thermodynamic limit of a bulk lattice. 
For each finite cluster, the Hamiltonian is decomposed in the usual
perturbative form $H = H_0 + \lambda V$, where $H_0$ describes a simple
system with known ground state and V is treated perturbatively to high order.
In the present work we use 'Ising expansions' in which $H_0$ consists of the
diagonal $S_i^zS_j^z$ terms, and V consists of the transverse quantum
fluctuations. Thus the SU(2) symmetry is broken by the choice of $H_0$,
which reflects the order in the chosen phase, but is restored in the limit
$\lambda = 1$. Provided there is no singularity for $0<\lambda<1$ the true 
ground state will be reached in this limit. We refer the readers to more
detailed expositions [12,13] for further details of the method.

\subsection{N\'eel phase}

To derive series to 8th order we need to consider clusters of up to 8 sites,
having 3 bond types. There are a total of 10,644 distinct such clusters  
for the 2D case. The ground state energy and magnetization are 
expressed in the form

\begin{equation}
                E_0/N = \sum_{n=0}^\infty a_n \lambda^n
\end{equation}

\begin{equation}
                  M  = \sum_{n=0}^\infty b_n \lambda^n
\end{equation}

\noindent with the coefficients $(a_n,b_n)$ computed to 12 figure accuracy.
The series are analysed by standard Pad\'e approximant methods to yield
estimates of $E_0$ and $M$ for any values of the exchange constants 
$(J_0,J_1,J_2)$. In the present work we choose $J_0=J_1=1$ and plot
quantities versus the frustration parameter $J_2/J_1$.

\subsection{2+2 phase}
The 2+2 phase has a 4-sublattice structure, and it is necessary to distinguish
two types of $J_1$ bond, between like and unlike spins. This results in a total
of 22,613 clusters with 4 bond types, to 8th order. The derivation and
analysis of the series then proceeds in the same way as above. 

\subsection{Spiral phase}
To carry out an Ising expansion for a non-collinear ordered phase we
transform to a local basis, in which each spin is directed along its local
z-axis. This results in a Hamiltonian of the form

\begin{equation}
 H = -{1\over 4} (J_0 + J_1 \cos{\theta} + J_2 cos{2\theta}) N + H_0 + \lambda V
\end{equation} 

\noindent with

\begin{eqnarray}
 H_0 &= &J_0 \sum_{<ij>}^{(0)} (-S_i^z S_j^z +{1\over 4})
 +J_1 cos\theta \sum_{<ij>}^{(1)} (-S_i^z S_j^z + {1\over 4}) \nonumber \\
   &  &  +J_2 cos{2\theta} \sum_{<ij>}^{(2)} (S_i^z S_j^z - {1\over 4})
\end{eqnarray}

\noindent and

\begin{eqnarray}
V & = & -{1\over 2}J_0 \sum_{<ij>}^{(0)}(S_i^+ S_j^+ + S_i^- S_j^-) \nonumber \\
  &   & -{1\over 4}J_1 \sum_{<ij>}^{(1)} [ (1 + cos{\theta}) 
         (S_i^+S_j^+ + S_i^-S_j^-) \nonumber \\ 
  &   &  -(1 - cos{\theta}) (S_i^+ S_j^- + S_i^- S_J^+ ) \nonumber \\
  &   &  +2sin{\theta}(S_i^+ S_J^z + S_i^- S_j^z - S_i^z S_j^+ - S_i^z S_j^-)]
\nonumber \\
  &   &  -{1\over 4}J_2 \sum_{<ij>}^{(2)} [ (1 - cos{2\theta}) (S_i^+ S_J^+  +
S_i^- S_J^-) \nonumber \\
  &   &  -(1+cos{2\theta}) (S_i^+ S_j^- + S_i^- S_j^+) \nonumber \\
  &   &  -2sin{2\theta}(S_i^+ S_j^z + S_i^ -S_j^z - S_i^z S_j^+ - S_i^z S_j^-)]
\end{eqnarray}

\noindent where the superscripts 0,1,2 refer to the 3 bond types, and $\theta$
is the angle between successive spins in columns (actually the angle is $\pi -
\theta$ in the original picture, before a rotation of axes).
This Hamiltonian contains the angle $\theta$ as a parameter, and this is not 
known a-priori. Thus we choose a range of values, plot the energy as a 
function of $\theta$, and choose the correct $\theta$ from the minimum. In
practice, the minimum is quite shallow and it is difficult to choose $\theta$
with high precision. However, this doed not seriously affect the energy
estimates.

\subsection{Results}

Figure 3 shows the ground state energy and magnetization versus $J_2/J_1$ 
for the 2D ANNNH model, for the N\'eel, spiral and 2+2 phases, obtained from
our series. The series have been analysed by standard Pad\'e approximant
techniques, using both the direct series and the logarithmic derivative.
The latter are found to give slightly more stable results, but the two
approaches are broadly consistent. Where error bars are shown in the figues,
they represent 'confidence limits' based on the spread of different
approximants.

We first comment on the ground state energy. These series are very regular,
and any uncertainty is estimated to be no larger than the plotted points.   
The N\'eel and spiral series appear to meet smoothly at a point near
$J_2 = 0.47 \pm 0.02$, well above the classical transition point $0.25$.
The series become a little erratic in the immediate vicinity of this value,
and the transition appears continuous. In Fig. 3a we also plot the energy
of the 2+2 phase. This clearly lies at higher energy, and is thus not a
stable phase. It seems that, as $J_2 \rightarrow \infty$, the energies 
of the spiral and 2+2 phases become asymptotically equal, as for the
classical case.

The magnetization series are less regular, and the error bars become 
quite large near the transition point. The most interesting feature is
that both the N\'eel and spiral phase magnetizations appear to be 
dropping to zero at the transition, between 0.45 and 0.5. Thus quantum
fluctuations in this 2D model are large enough to destroy the long-
range order at this point. Indeed we cannot exclude the possibility of
a (very) narrow non-magnetic phase. We also show the magnetization for
the 2+2 phase, but, since this phase has higher energy, it is of little
significance.

\begin{figure}
 \includegraphics[width=7cm]{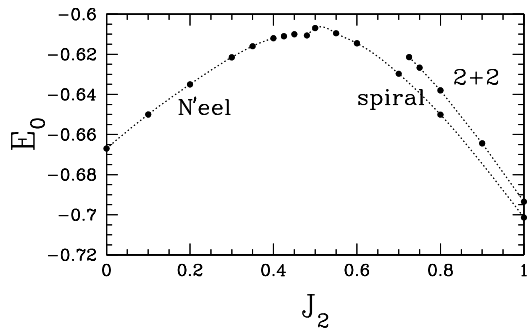}
 \vspace{0.5cm}
 \includegraphics[width=7cm]{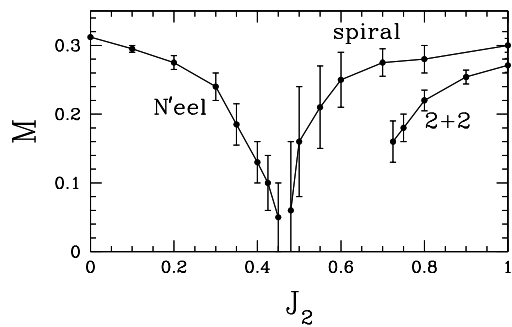}
 \caption{Ground state energy (upper panel) and magnetization (lower panel)
as a function of $J_2$ for the 2D ANNNH model}
 \label{fig3}
\end{figure}

\section{HIGH T SERIES FOR S(q)}
\label{sec3}

High temperature series [12] provide a complementary approach for studying
the nature of magnetic orders. Although the Mermin-Wagner theorem
precludes any finite temperature ordered phase in the 2D model, it is
expected that, as the temperature is lowered, the correlations that build up 
will reflect the nature of the order which occurs at T=0.
High T expansions for a correlator $<S_i^z S_j^z>$ can be developed as

\begin{equation}
       C({\bf r}) = {1\over Z} \sum_{n=0}^\infty {(-1)^n\over n\\!}
 Tr\{[S_0^z S_{\bf r}^z H^n \} \beta^n
\end{equation}

\noindent where $\beta = 1/k_B T$, and $Z$ is the partition function, 
which is itself expanded as a series in $\beta$. We note that, since we are in
a paramagnetic phase, the correlations have full rotational symmetry
and it suffices to compute the (zz) correlators.

From these we compute a high T series, in powers of $\beta$, for the static
structure factor

\begin{equation}
            S({\bf q}) = \sum_{\bf r} e^{i {\bf q \cdot r}}  C({\bf r})
\end{equation}

\noindent which should develop a peak at whatever $q$ value reflects the 
T = 0 order.

To compute the structure factor series to 8th order, for general $\bf q$,
would require correlator series for all cluster space-types with 8 or fewer
bonds, a total of over 600,000 distinct clusters. However, for $\bf q$ in the
modulation direction, this number can be reduced considerably, by effectively
calculating correlator series between horizontal rows of spins. This requires
only 76712 clusters.

The $S({\bf q})$ series converge rapidly at high T (small $\beta$) and can be 
evaluated using Pad\'e approximants down to about $t=k_B T/J_1 \sim 0.5$.
We have carried out such an analysis for
$\bf q = (\pi,q_z)$ for various $J_2$ and results are shown in Fig. 4, for
the temperature $t=0.5$. Below this $t$ the series become too erratic. We
see that for $J_2 = 1.0$ there is a rather sharp peak at $q_z=0.58 \pi$, 
corresponding to an angle of 76 degrees. As $J_2$ is decreased, the peak
broadens and moves to larger $q_z$ (smaller angles). Note that $q_z=\pi/2$ 
would correspond to a modulation wavelength of a lattice spacings, as for the
2+2 structure, whereas $q=\pi$ corresponds to the N'eel phase. The peak positions
do not change significantly with temperature.

\begin{figure}
 \includegraphics[width=0.8\linewidth]{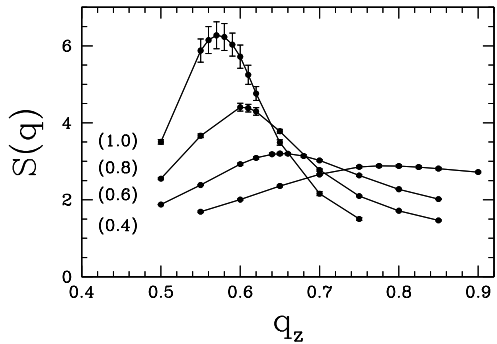}
 \caption{Structure factor $S(q)$ with q in units of $\pi/a$, as calculated
from high T series expansions}
 \label{fig4}
\end{figure}

\section{THE 3D ANNNH MODEL}
\label{sec5}

We have used the same approach to study the ANNNH model on the simple-
cubic lattice.

The ground state energy and magnetization are shown in Figure 5.

\begin{figure}
 \includegraphics[width=7cm]{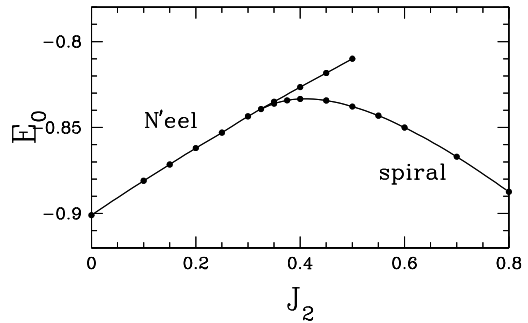}
 \vspace{0.5cm}
 \includegraphics[width=7cm]{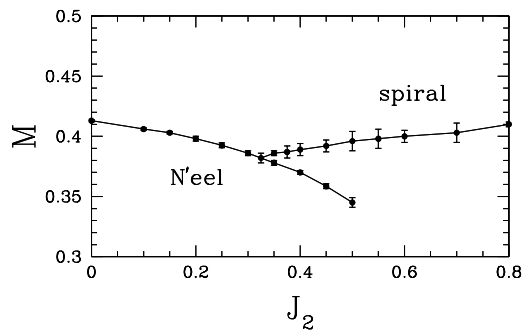}
 \caption{Ground state energy (upper panel) and magnetization (lower
panel) for the 3D ANNNH model, from series expansions.}
 \label{fig5}
\end{figure}

The following points can be noted: \newline
1. The energy series are very regular, and the curves meet smoothly at
$J_2 = 0.34 \pm 0.01$. The N\'eel series can be accurately continued well
beyond this point, as shown. We also note the maximum in the spiral phase
energy near 0.4, which then drops again to meet the N\'eel curve smoothly.
This feature occurs in the classical case, and is also apparent, though
less clearly, in the 2D case (Fig.3). The /N\'eel/spiral crossover point,
at $J_2 \sim 0.34$, is again well above the classical value 0.25, but
the difference is less than in the 2D case, reflecting the smaller 
quantum fluctuations in higher dimension. \newline
2. The magnetization series are less regular, and this is reflected in
the error bars, although the size of the uncertainty is exaggerated by
the scale chosen for the figure. We note that the magnetization
decreases on approaching the crossover point from either side, but only
by approx. 10
zero, again showing that quantum fluctuations are less dominant.

As in the 2D case, we have also derived high T series for the structure  
factor $S(q)$. There is, however, one important difference. In 3D the system
will retain long-range magnetic order at finite temperature, up to some 
critical temperature $T_c (J_2)$. On approaching $T_c (J_2)$ from above,
the structure factor $S(q)$ at the appropriate wavevector is expected to        diverge in the thermodynamic limit, reflecting the development of long-
range correlations at the critical temperature. Thus we may expect to be 
able to estimate the locus of this critical line from our series. Some
results are shown in Figure 6.

\begin{figure}
 \includegraphics[width=7.0cm]{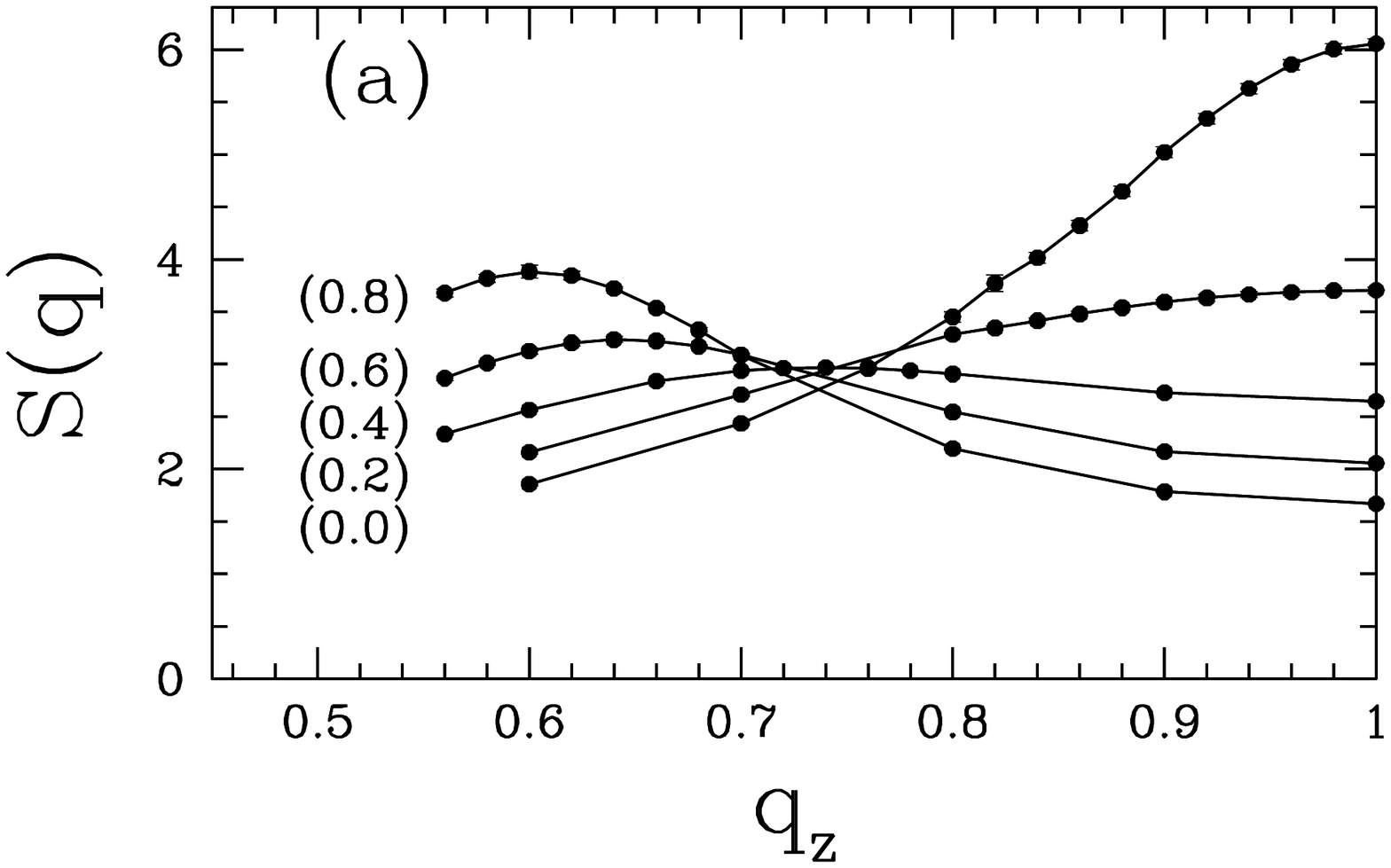}
 \includegraphics[width=7.0cm]{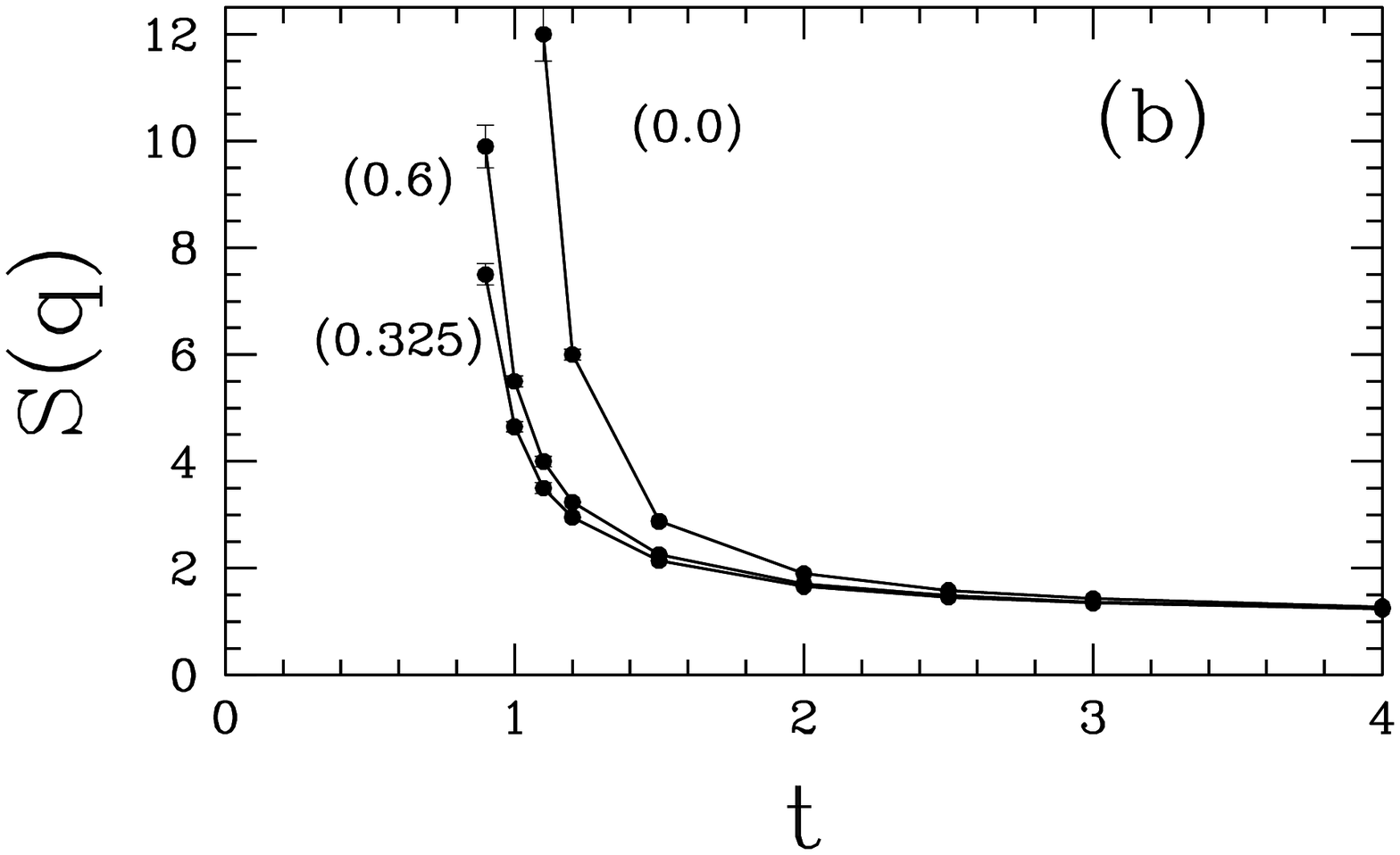}
 \caption{(a)Structure factor $S(q)$ versus $q_z$ at a temperature t=1.2,  
for several values of $J_2$, for the 3D ANNNH model; (b) structure factor
at the critical wavevector versus temperature, for several values of $J_2$.}
 \label{fig6}
\end{figure}

Figure 6(a) shows $S(q)$ versus $q_z$ for various $J_2$. For $J_2=0.0,0.2$ the 
maximum is at $q_z=1.0$, corresponding to N\'eel order. For larger $J_2$ the
peak moves continuously to smaller $q_z$: $q_z \tilde 0.73$ for $J_2=0.4$,
$q_z \tilde 0.64$ for $J_2=0.6$, $q_z \tilde 0.6$ for $J_2=0.8$. This is
indicative of an incommensurate spiral phase. The $q_z$ values are consistent
with those found to give the lowest ground-state energy in the T=0 spiral
phase. The point where the peak begins to move away from $q_z=1$ is close to
$J_2=0.325$ .

We have also analysed the $S(q)$ series to estimate the values of the critical
temperature as a function of $J_2$. While the 8th order series are too short to
provide estimated of high precision, P\'ade approximants to the logarithmic
derivative series do show a fairly consistent pole on the positive real axis,
corresponding to a line of critical points.

In Figure 6(b) we plot the value of $S(q_c)$, at the critical wavevector $q_c$,
versus reduced temperature $t=k_B T_c /J_0$. A strong divergence is clearly 
seen. Our best estimates for the critical temperatures are: 1.09 ($J_2=0.0$),
0.82 ($J_2=0.325$), 0.89 ($J_2=0.6$). For $J_2=0$, the isotropic simple-cubic
nearest-neighbor model, a more precise estimate is available from
longer series [14]. There are, as far as we know, no previous estimates of the
critical line for the 3D annnh model. The critical temperature is lowest near
$J_2=0.324$, which is also where the peak in $S(q)$ moves away from $q_z=1$.
This is a Lifshitz point [15], where paramagnetic, N\'eel ordered and spiral
phases meet and coexist.

\section{SUMMARY AND DISCUSSION}
\label{sec6}

We have used a combination of perturbation series at T=0 and high T
expansions to investigate the nature of magnetic order, and the magnetic 
phase diagram in the quantum spin S=1/2 ANNNH model, in both 2 and 3 dimensions.
While it is easy to show, for classical vector spins, that an incommensurate
spiral phase exists for large frustration $J_2$, previous analytic studies
for quantum spins have encountered difficulties. Our study confirms that the
classical picture remains qualitatively correct. However, quantum fluctuations
shift the crossover point between N\'eel and spiral phases substantially.

For the 2D model we find that the magnetizations in both N\'eel and spiral
ground states appear to tend continuously to zero at the crossover point. 
This was not expected, and is reminiscent of the behaviour in the   
$J_1-J_2$ model, where there is an intermediate non-magnetic phase.
We see no evidence for such a phase here, although we cannot exclude the
possibility of a very narrow phase of this kind. In the 3D model, the 
magnetizations clearly cross over at a finite value.

In the 3D model, the magnetic phases extend to finite temperature, and we
have estimated the position of the critical line, and of the Lifshitz
point, where paramagnetic, N\'eel and spiral phases coexist.

In the large $J_2$ limit a collinear phase, the '2+2 phase', becomes 
asymptotically degenerate with the $q_z = \pi/2$ spiral, both having a
modulation wavelength of 4 lattice spacings. Such a phase, termed 'pair-
checkerboard', was claimed to exist in the FeSe monolayer system [11].
We find that, in the ANNNH model, such a phase always has higher energy
than the spiral. Thus, if it exists as a stable phase, a more complex
Hamiltonian would be indicated.

\begin{acknowledgments}
This work is supported in part by NSF grant number DMR-1306048, and by
computing resources provided by the Australian (APAC) National Facility.
\end{acknowledgments}

\end{document}